\documentclass{elsart}
\usepackage[english]{babel}
\usepackage{epsfig}

\newcommand{\hb} {HERA-B}
\newcommand{\ra} {\mbox{$\mskip 3mu \rightarrow \mskip 5mu$}}
\newcommand{\mt} {\mbox{$p_T$}}
\newcommand{\xeff} {\mbox{$x_F$}}
\newcommand{\sqs} {\mbox{$\sqrt{s}$}}
\newcommand{\JPsi}{\mbox{\ensuremath{J/\psi}}}
\newcommand{\Psizs}{\mbox{\ensuremath{\psi(2S)}}} 
\newcommand{\JPsimm} {\mbox{\ensuremath{J/\psi \rightarrow \mu^+\mu^-}}}  %  J/psi 
\newcommand{\JPsiee} {\mbox{\ensuremath{J/\psi \rightarrow \mathrm{e}^+ \mathrm{e}^-}}}  %  J/psi
\newcommand{\JPsill} {\mbox{\ensuremath{J/\psi \rightarrow \ell^+ \ell^-}}}  %  J/psi
\newcommand{\bbbar}{\mbox{\ensuremath{b\overline{b}}}}         % b bbar
\newcommand{\dilepton}{\mbox{\ensuremath{\ell^+ \ell^-}}}               % lepton pair
\newcommand{\epem}{\mbox{\ensuremath{e^+ e^-}}}       % e+ e-
\newcommand{\mpmm}{\mbox{\ensuremath{\mu^+ \mu^-}}}                     % mu+ mu-

\begin{document}

\begin{frontmatter}

\title{Measurement of the $\mathbf{J/\psi}$ Production Cross Section in
920 GeV/c Fixed-Target Proton-Nucleus Interactions}
\vspace*{0.2cm}
\centerline{\bf{HERA-B Collaboration}}
\vspace*{0.2cm}
I.~Abt$^{23}$,
M.~Adams$^{10}$,
M.~Agari$^{13}$,
H.~Albrecht$^{12}$,
A.~Aleksandrov$^{29}$,
V.~Amaral$^{8}$,
A.~Amorim$^{8}$,
S.~J.~Aplin$^{12}$,
V.~Aushev$^{16}$,
Y.~Bagaturia$^{12,36}$
% ...

\vspace{5mm}
\noindent
% ...
$^{8}${\it LIP Coimbra, P-3004-516 Coimbra,  Portugal~$^{f}$} \\
% ...
$^{10}${\it Institut f\"ur Physik, Universit\"at Dortmund, D-44221 Dortmund, Germany~$^{d}$} \\
% ...
$^{12}${\it DESY, D-22603 Hamburg, Germany} \\
$^{13}${\it Max-Planck-Institut f\"ur Kernphysik, D-69117 Heidelberg, Germany~$^{d}$} \\
% ...
$^{16}${\it Institute for Nuclear Research, Ukrainian Academy of Science, 03680 Kiev, Ukraine~$^{h}$} \\
% ...
$^{23}${\it Max-Planck-Institut f\"ur Physik, Werner-Heisenberg-Institut, D-80805 M\"unchen, Germany~$^{d}$} \\
% ...
$^{29}${\it DESY, D-15738 Zeuthen, Germany} \\
% ...
$^{36}${\it visitor from High Energy Physics Institute, 380086 Tbilisi, Georgia} \\
% ...
$^{d}$ supported by the Bundesministerium f\"ur Bildung und Forschung, FRG, under contract numbers 05-7BU35I, 05-7DO55P, 05-HB1HRA, 05-HB1KHA, 05-HB1PEA, 05-HB1PSA, 05-HB1VHA, 05-HB9HRA, 05-7HD15I, 05-7MP25I, 05-7SI75I \\
% ...
$^{f}$ supported by the Portuguese Funda\c c\~ao para a Ci\^encia e Tecnologia under the program POCTI \\
% ...
$^{h}$ supported by the National Academy of Science and the Ministry of Education and Science of Ukraine \\
% ...

\begin{abstract}
The mid-rapidity ($d\sigma_{pN}/dy$ at $y$=0) and total ($\sigma_{pN}$) production
cross sections of \JPsi\ mesons are measured in proton-nucleus interactions. Data
collected by the \hb\ experiment in interactions of 920 GeV/c protons with carbon,
titanium and tungsten targets are used for this analysis. The \JPsi\ mesons are
reconstructed by their decay into lepton pairs. The total production cross section
obtained is $\sigma_{pN}^{J / \psi} =  663 \pm 74 \pm 46$ nb/nucleon.
In addition, our result is compared with previous measurements.
\\
\vspace*{0.2cm}
\end{abstract}

\end{frontmatter}

\section{Introduction}
\label{intro}
Since the spectacular discovery of the \JPsi\ particle~\cite{Aubert,Augustin}
heavy-quarkonium
production in hadron-hadron interactions has been in the focus of interest
because it provides important information on both perturbative and non-perturbative
aspects of Quantum Chromodynamics (QCD). Especially the unexpectedly large cross sections
for \JPsi\ and \Psizs\ at large transverse momenta observed by the CDF
experiment~\cite{CDF} renewed this interest and led to the development of the
non-relativistic QCD (NRQCD) approach~\cite{Bodwin}, which extends the color-singlet
model by including
color-octet contributions~\cite{Cho,Braaten,Beneke}.
A different approach to charmonium
production is based on the color evaporation model~\cite{Amundson,Ingelman}. Both
models predict the energy dependence of \JPsi\ production.

These theoretical predictions have to be compared to the experimental results obtained
in proton induced reactions. However, despite the large interest in this field,
the experimental situation is far from being satisfactory. Cross section measurements
at comparable energies differ well outside the quoted uncertainties. Thus a new measurement
with low systematic errors is of interest.

In this paper, we report on a measurement of the cross section  for \JPsi\
production in interactions of 920 GeV/c protons with nuclei of atomic weight A
using the HERA-B detector. The corresponding center of mass energy of the
proton nucleon interaction is \sqs =41.6~GeV.
A data sample
of 210~million interactions on carbon, titanium and tungsten targets was
recorded using a minimum bias trigger in the 2002-2003 HERA running period.
The \JPsi\ mesons are detected in the inclusive reaction
\begin{displaymath}
   p A \ra\ \JPsi\ X \: \rm{with} \; \JPsi\ \ra\ \epem\ \, \rm{or} \;\, \mpmm\ .
\end{displaymath}
The advantage of such a minimum bias data sample is the low systematic
error due to the large ($>97\%$) trigger efficiency, the large angular coverage
of HERA-B and its large reconstruction efficiency. On the other hand, the size
of the sample is small, since charmonium cross sections
are $4-5$ orders of magnitude less than the total inelastic cross section.

\section{Apparatus}
\label{Appa}

\begin{figure}[htb]
\begin{center}
\epsfig{file=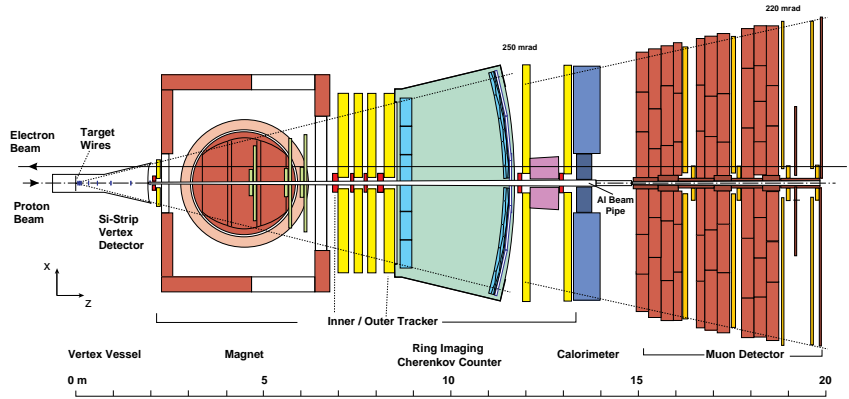,width=0.80\columnwidth}
 \caption{\small Plan view (bending plane) of the HERA-B detector.
    \label{fig:layout} }
\end{center}
\end{figure}

The HERA-B fixed-target spectrometer operated at the 920 GeV/c proton
beam of the HERA storage ring at DESY using one or more wire
targets inserted into the beam halo. The detector was equipped with a vertex
detector and extensive tracking and particle identification systems. It
had a large geometrical coverage from 15 mrad to 220 mrad in the
bending (horizontal) plane and 15 mrad to 160 mrad in the non-bending
(vertical) plane. Fig.~\ref{fig:layout}
shows a plan view of the detector in the configuration of the 2002-2003
data run.

The target system~\cite{ehr00} consisted of two stations of four wires
each. The wires were positioned above, below, and on either side of the
beam and were made from various materials including carbon, titanium
and tungsten. The stations were separated by 40~mm along the beam
direction.  The wires were positioned individually in the halo of
the stored proton beam and the interaction rate for each inserted wire
was adjusted independently.

The Vertex Detector System (VDS)~\cite{bau03} was a forward micro-strip vertex
detector integrated into the HERA proton ring. It provided a precise
measurement of primary and secondary vertices. The VDS consisted of 7
stations (with 4 stereo views each) of double-sided silicon strip detectors
($50\,\times\,70$\,mm, 50 $\mu$m pitch) integrated into a Roman pot system inside a
vacuum vessel and operated as near as 10\,mm from the beam. An additional
station was mounted immediately downstream of the 3\,mm thick
aluminum window of the vacuum vessel.

The first station of the main tracker was placed upstream of the
2.13\,Tm spectrometer dipole magnet. The remaining 6 tracking
stations extended from the downstream end of the magnet to the
electromagnetic calorimeter (ECAL) located 13\,m downstream of the target.
Each tracking station was divided into inner and outer detectors.  
The region starting from the beam pipe and extending up to 200\,mm was
covered by micro-strip gas chambers with GEM foils (inner tracker~\cite{gra01})
which, however, were not used in this analysis due to insufficient stability.
The region outside the inner tracker was covered by the large area outer
tracker (OTR)~\cite{alb05} consisting of $\approx$95,000 channels
of honeycomb drift cells.

Particle identification was performed by a Ring Imaging Cherenkov
detector (RICH)~\cite{ari04}, an electromagnetic calorimeter
(ECAL)~\cite{avo01} and a muon detector (MUON)~\cite{eig01}.
The RICH used $\mathrm{C_4F_{10}}$ as radiator gas and two large spherical mirrors
to project Cherenkov photons on the photon detector employing
multi-anode photomultipliers. The ECAL was based on
``shashlik'' sampling calorimeter technology, consisting of scintillator layers
sandwiched between metal absorbers. It was subdivided in three different sections
with increasing cell size. In the radially innermost section,
tungsten was used as an absorber, and lead was used everywhere else. The MUON
detector was segmented into four super-layers. Iron and concrete shielding
extended from just behind the ECAL to the last MUON super-layer, except for
gaps for the super-layers themselves.
The first two super-layers consisted of three layers of tube chambers with
different stereo angles. The last two super-layers each consisted of one layer
of tube chambers with additional cathode pad readout.

\section{Data Sample and Trigger}
\label{DataTri}

The present analysis is performed on single wire carbon, titanium and
tungsten runs taken under stable conditions with a minimum bias trigger 
with a total of 182~million interactions (Table~\ref{tab:numevents}). 
The trigger required at least 20
hits in the RICH detector (compared to an average of 33 for a full
ring from a $\beta=1$ particle~\cite{ari04}) or an energy deposit
of at least 1~GeV
in the electromagnetic calorimeter and was sensitive to
$\epsilon_{trigger} > 97\%$ of the
total inelastic cross section $\sigma_{inel}$.

The integrated luminosity was determined~\cite{lumi} from the number of
inelastic interactions $N_{inel}$ using the expression
$\mathcal{L} = N_{inel} / (\epsilon_{trigger} \cdot \sigma_{inel})$.
The data were recorded at a moderate interaction rate of about 1.5 MHz
which corresponds to 0.17 interactions per filled bunch crossing. Therefore
only about $10\%$ of the events contain more than one interaction.
The high data acquisition rate of about 1000 Hz allowed to record the
bulk of the data within two weeks.
 
The information delivered by the vertex detector, outer tracker,
electromagnetic calorimeter and muon detector entered into this
analysis. 

\begin{table}[ht]
\begin{center}
\begin{tabular}{|c|c|c|c|}
\hline
Target & Events & $\mathcal{L}(\mu b^{-1})$ & $\Delta (\mathcal{L}) / \mathcal{L}$
\\ \hline \hline
C Below 1  & 68.8 M & 277.  & 4.8$\%$
\\ \hline
C Inner 2  & 20.5 M &  98.0 & 4.8$\%$
\\ \hline 
Ti Below 2 & 24.7 M &  30.9 & 5.0$\%$
\\ \hline
W Inner 1  & 67.6 M &  35.7 & 3.8$\%$
\\ \hline 
\end{tabular}
\vspace*{0.3cm}
\caption{Summary of minimum bias statistics. Target material, wire position
with respect to the proton beam 
(inner wire corresponds to +X and outer wire to -X, see Fig.~\ref{fig:layout}),
integrated luminosities as well as the sample-specific luminosity errors
are given. In addition, a 2.0$\%$ scaling error on luminosity has to
be taken into account.
\label{tab:numevents}}
\end{center}
\end{table}

\section{Data Analysis}
\label{DataAn}
\subsection{Event Selection}

The search for \JPsi\ candidates is performed by analyzing the invariant mass
spectrum of unlike-sign lepton pairs. The phase-space region of the
\JPsi , in rapidity $y$, covered by our measurement is $-1.25<y<+0.35$.
The corresponding range in Feynman \xeff\  is $-0.225<x_F<+0.075$.
Throughout this paper, the rapidity is defined in the proton-nucleon
center-of-mass system.
Soft cuts (see Table~\ref{tab:presel}) are applied on the number of hits
in the tracking system to select properly reconstructed tracks. Cutting on the
dilepton vertex fit probability ensures that the two tracks have a common vertex.
A cut on the transverse momentum of the leptons helps to reduce the background
with a small loss ($< 3\%$) in signal efficiency.

\begin{table*}[htb]
\begin{center}
\begin{tabular}{|l|p{3.0cm}|p{3.0cm}|}
\hline
              &  \multicolumn{1}{c|}{\mpmm\ } & \multicolumn{1}{c|}{\epem\ }
\\ \hline \hline
Number of hits in VDS & \multicolumn{2}{c|}{$\geq$ 6}  \\
(12 hits on average)  & \multicolumn{2}{c|}{ }
\\ \hline
Number of hits in OTR & \multicolumn{2}{c|}{$\geq$ 10} \\
(40 hits on average)  & \multicolumn{2}{c|}{ }
\\ \hline
Transverse mom. \mt\  (GeV/c) & \multicolumn{2}{c|}{$>$ 0.7}
\\ \hline  
Dilepton rapidity & \multicolumn{2}{c|}{$-1.25<y<+0.35$} 
\\ \hline
Dilepton vertex fit probability & \multicolumn{2}{c|}{$> 10^{-5}$}
\\ \hline \hline
Muon likelihood ($L_\mu$)  & \multicolumn{1}{c|}{$> 0.2$} & 
\\ \hline
VDS-ECAL match                  &   & $|\Delta Y| < 0.75$ cell width   
\\ \hline
$E/p$                           &   & $-2.0< (E/p-0.98)/\sigma <+3.5 $   
\\ \hline
\end{tabular}
\vspace*{0.3cm}
\caption{Event selection requirements. The lepton identification cuts
(last three rows) are fixed by the optimization procedure described
in the text. \label{tab:presel}}
\end{center}
\end{table*}

The signal to background ratio can be improved by applying more stringent
cuts on the lepton identification. In order to find the best set of cut values
the signal significance $ S / \sqrt{S+B}$ is maximized. S is the
number of \JPsi\ surviving the selection cuts in the \JPsi\ Monte Carlo
sample (see Sec.~\ref{MC}) scaled to obtain a comparable number
of reconstructed \JPsi\  in Monte Carlo as in real data.
The background B is evaluated from the like sign invariant mass spectrum
in a 2.5 standard deviation window around the \JPsi\  position.
Such a procedure allows to adjust the lepton identification criteria
in an unbiased way.

Muon candidates are selected by requiring that the muon likelihood $L_\mu$
(normalized to one) derived from the muon hit information is greater
than 0.2. 

In the case of electrons, the background is higher due to pions interacting in
the ECAL and charged hadrons overlapping with energy depositions by neutral particles.
We therefore
require that the track position extrapolated to the ECAL in the non-bending
direction Y matches a cluster with $|\Delta$Y$|$ $<$ 0.75 cell width.
Moreover, the ratio $E/p$, where $E$ is the energy deposited in the
electromagnetic calorimeter and $p$ the momentum
measured by the tracking system, provides good discrimination between electrons
and hadrons. However, the electrons can emit Bremsstrahlung photons (BR)
when passing the detector planes.
BR photons emitted before the magnet are searched for in the ECAL and the momentum of
the electron is corrected if such a cluster is found. Details of this procedure
are explained in~\cite{bbbar}.

Additional energy losses
result in a small deviation of the ratio $E/p$ from unity. Consequently, a
track is identified as an electron candidate if $-2.0 < (E/p - 0.98) / \sigma < +3.5$
with $\sigma \approx 6\%$.

\subsection{\JPsimm }

\begin{figure}[thb]
\begin{center}
\epsfig{figure=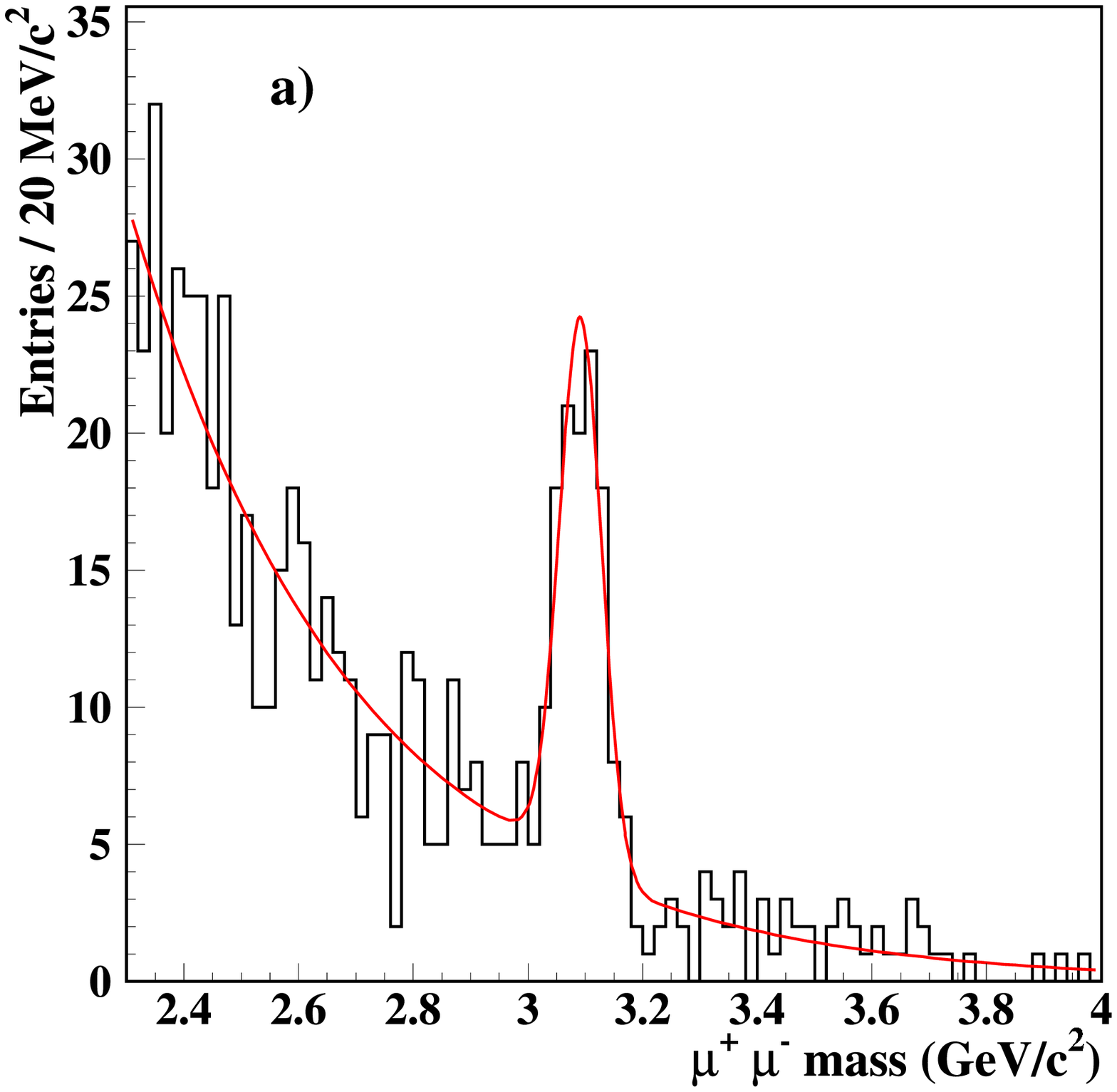,width=0.43\columnwidth}
\epsfig{figure=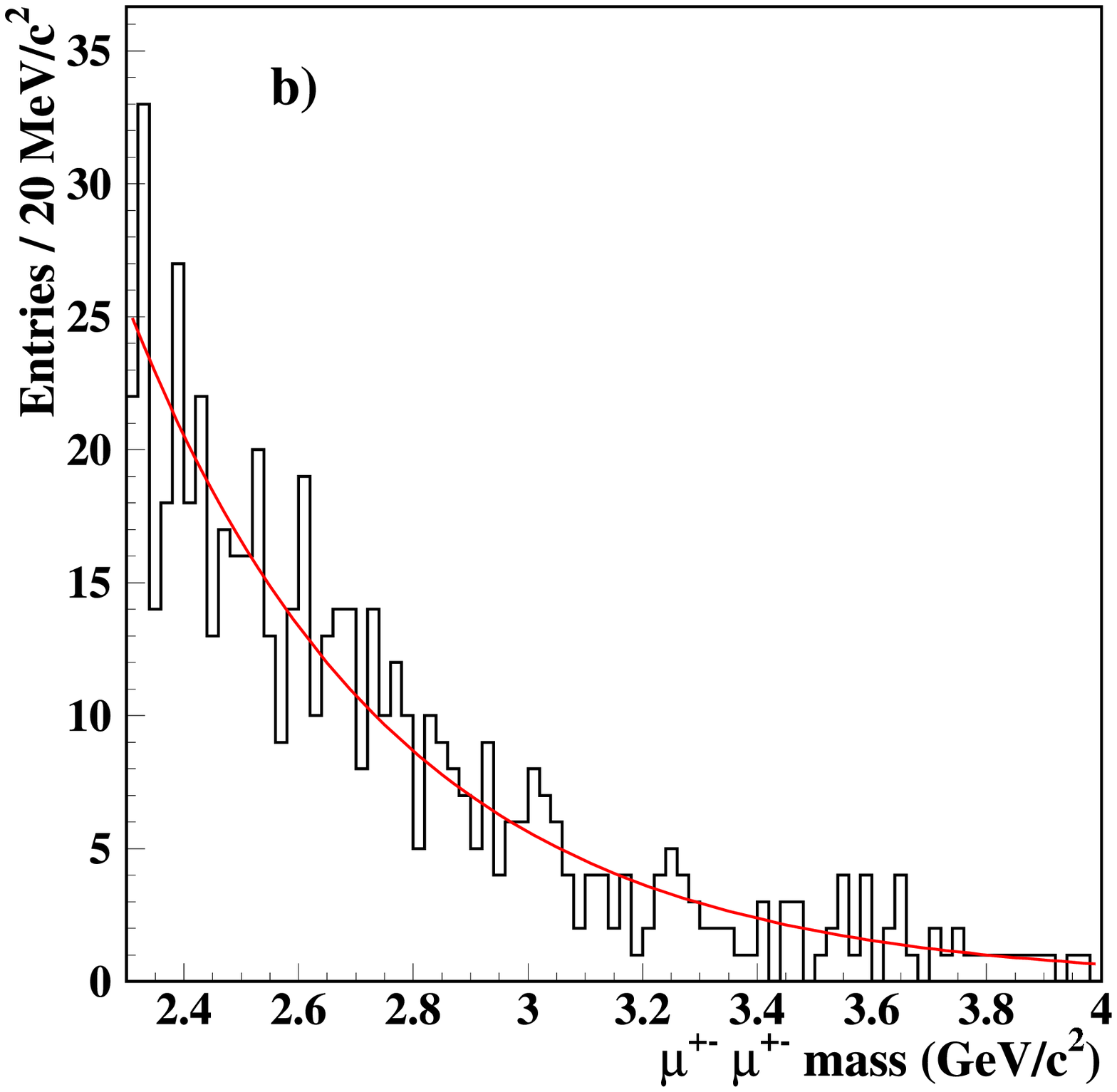,width=0.43\columnwidth}
\vspace*{-1.3cm}
\caption{a) Unlike sign pairs fitted with a Gaussian plus
a radiative tail for the signal and an exponential for the background;
b) like sign dimuon candidates fitted with an exponential.
\label{fig:mumusel}}
\end{center}
\end{figure}

The dimuon invariant mass distribution that results from combining all
data-sets and applying the selection criteria discussed above is
plotted in Fig.~\ref{fig:mumusel}a. This distribution is fitted by a
Gaussian plus a tail in the signal region and an exponential for the
background. The tail takes into account the radiative process
{\ensuremath{\mathrm{J}/\psi \rightarrow \mu^+\mu^- \gamma}}~\cite{Spiridonov}
whereas the background shape is suggested by the
mass spectrum of like-sign lepton pairs (see Fig.~\ref{fig:mumusel}b).  

A fit with this function
results in $100\pm12$ \JPsimm\ decays. To indicate the influence of the
radiative tail, the mass distribution is fitted with a simple Gaussian
resulting in $94\pm12$ events.

\begin{figure}[ht]
\begin{center}
\epsfig{figure=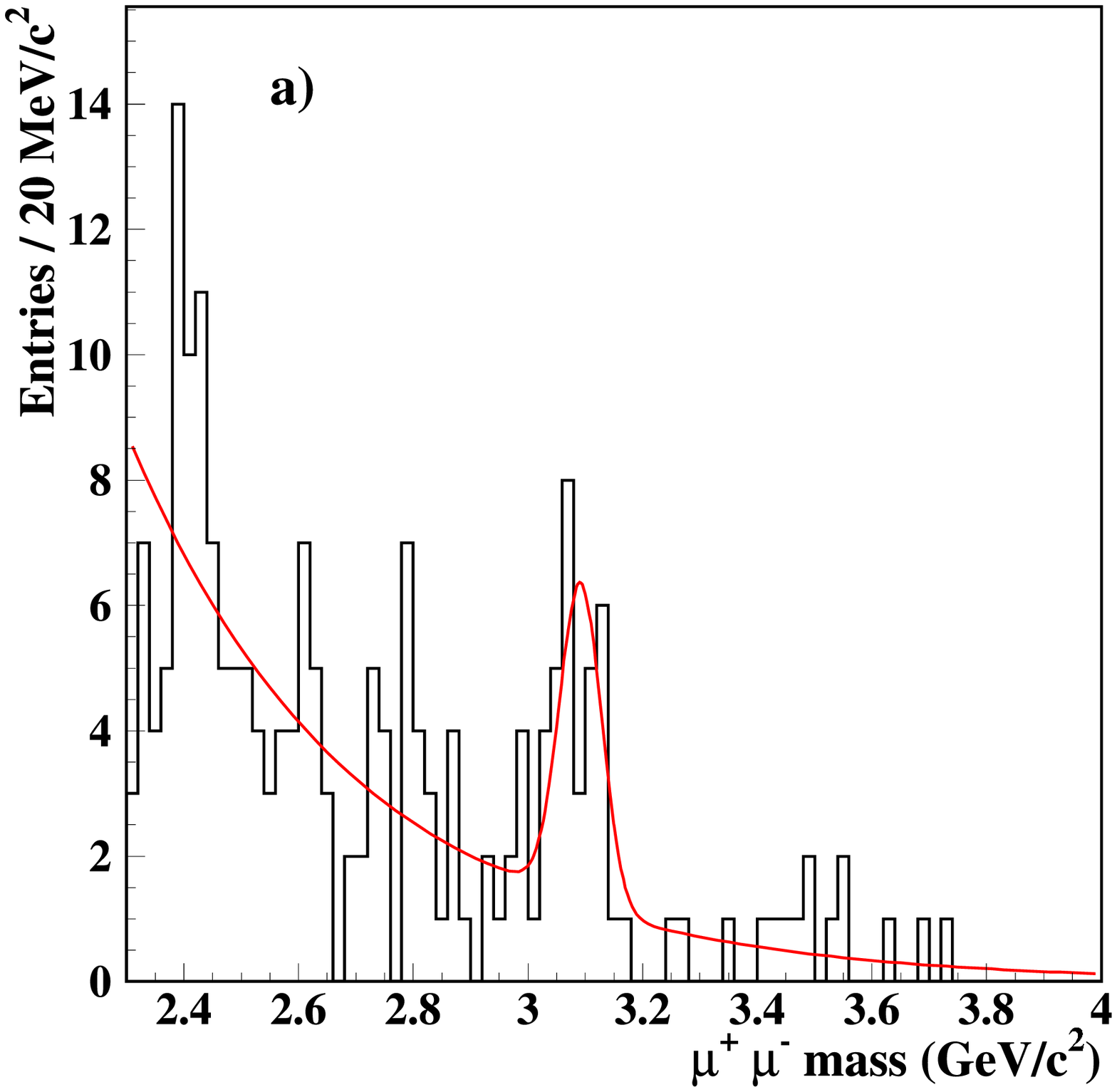,width=0.43\columnwidth}
\epsfig{figure=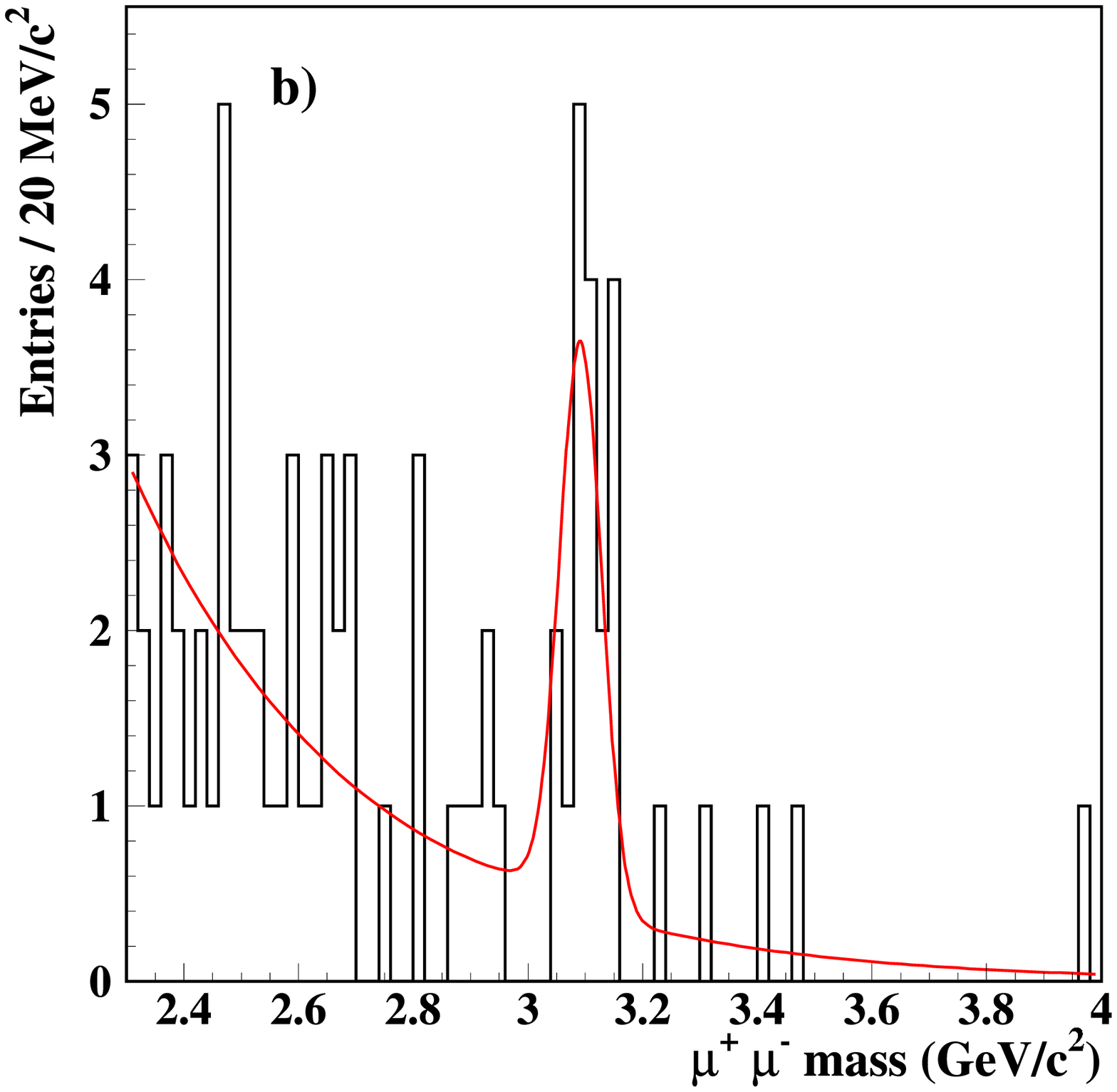,width=0.43\columnwidth}
\epsfig{figure=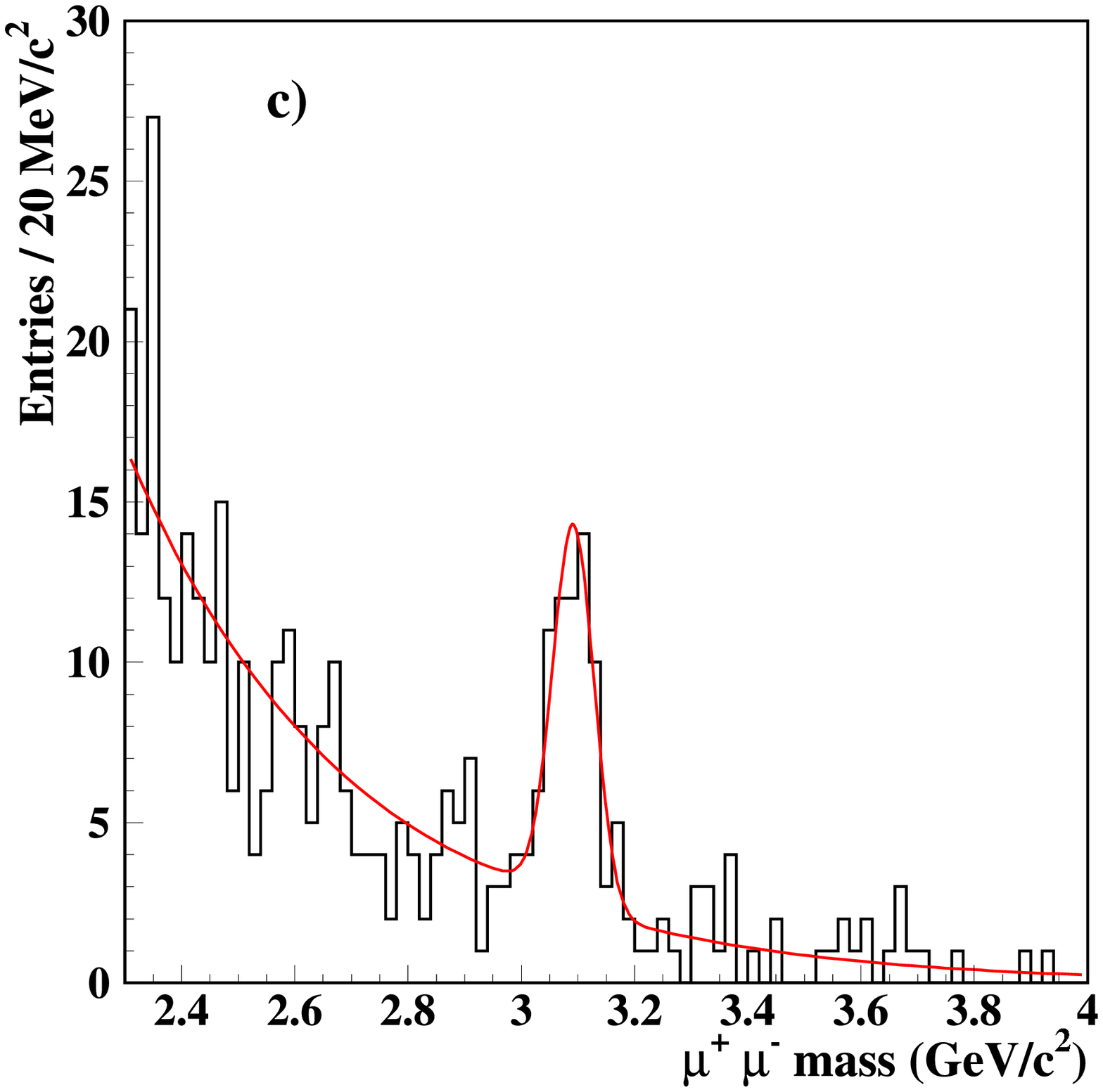,width=0.43\columnwidth}
\vspace*{-1.3cm}
\caption{Invariant mass of unlike sign muon pairs
from a) carbon, b) titanium, and c) tungsten sample.
\label{fig:mumuysel}}
\end{center}
\end{figure}

The peak position and the width (FWHM) are
$3.093 \pm 0.005$~GeV/$\mathrm{c^2}$ and $85 \pm 10$~MeV/$\mathrm{c^2}$, respectively.
The width is in good agreement with the width expected from the
\JPsi\ Monte Carlo simulation (see Sec.~\ref{MC}).
These fit results are used as input to the fit of the individual signals obtained
from the three different data samples shown in Fig.~\ref{fig:mumuysel}a-c.
The sum of the three signals (see Table~\ref{tab:NumJPsi}) is in good
agreement with the fit to the total sample.

\begin{table}[ht]
\begin{center}
\begin{tabular}{|l|r|r|}
\hline
\boldmath{\JPsimm\  }        & total sample & subsample 
\\ \hline \hline
Fit including radiative tail & 100$\pm$12 &
\\ \hline
\hspace*{1.0cm} carbon       &             & 25$\pm$6
\\ \hline
\hspace*{1.0cm} titanium     &             & 16$\pm$4
\\ \hline
\hspace*{1.0cm} tungsten     &             & 58$\pm$9
\\ \hline \hline
\boldmath{\JPsiee\  }        & total sample & subsample 
\\ \hline \hline
Fit including radiative tail & 57$\pm$13 &
\\ \hline
\hspace*{1.0cm} carbon       &             & 32$\pm$8
\\ \hline
\hspace*{1.0cm} titanium     &             &  2$\pm$4
\\ \hline
\hspace*{1.0cm} tungsten     &             & 24$\pm$10
\\ \hline
\end{tabular}
\vspace*{0.3cm}
\caption{Signal yield for \JPsimm\ and \JPsiee\ . \label{tab:NumJPsi}}
\end{center}
\end{table}

\subsection{\JPsiee }

\begin{figure}[thb]
\begin{center}
\epsfig{figure=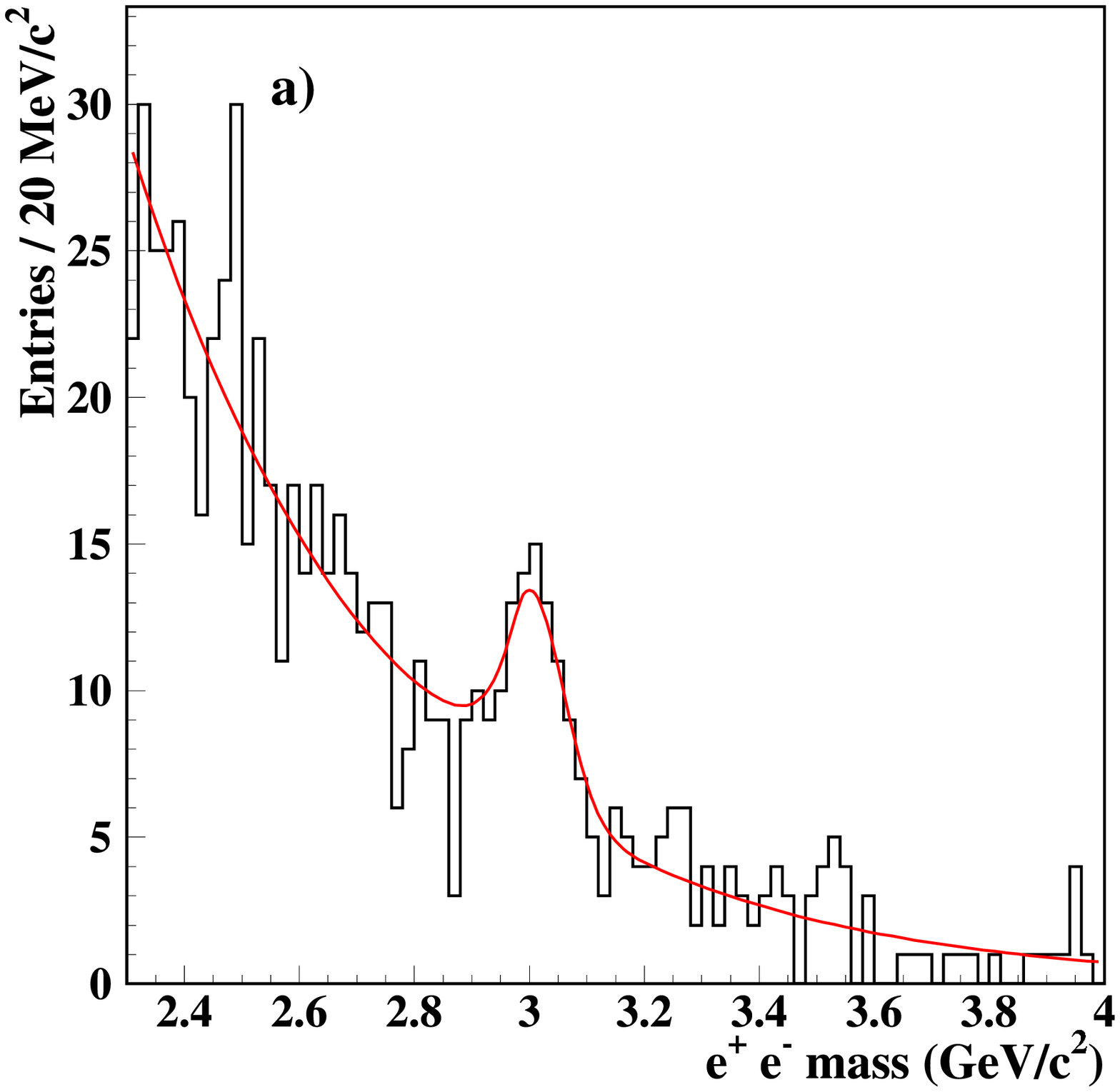,width=0.43\columnwidth}
\epsfig{figure=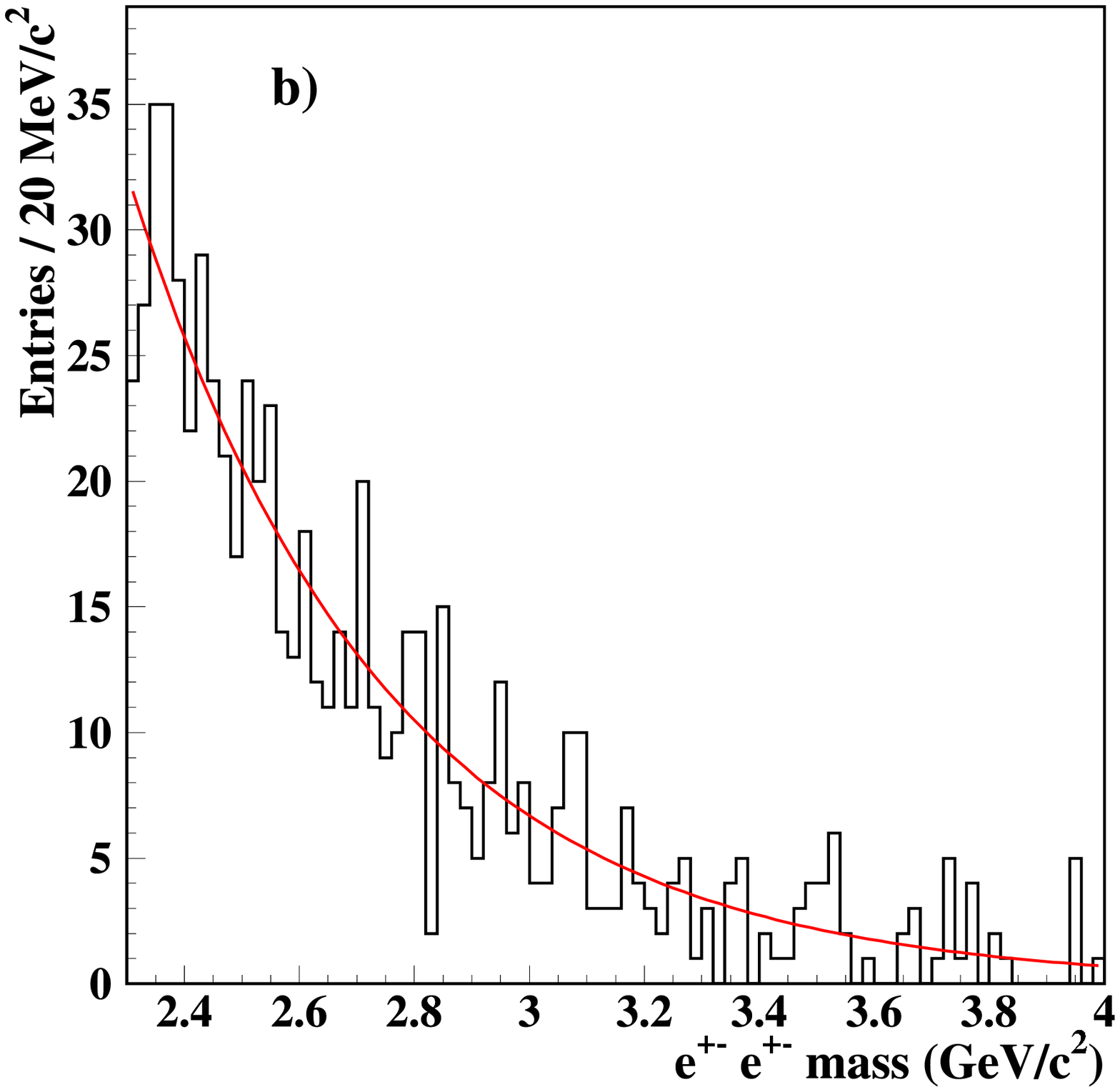,width=0.43\columnwidth}
\vspace*{-1.3cm}
\caption{a) Invariant mass of unlike sign electron pairs fitted with a Gaussian
plus a radiative tail for the signal and an exponential for the background;
b) like sign dielectron candidates fitted with an exponential.
\label{fig:eesel}}
\end{center}
\end{figure}

In contrast to muons, the momentum measurement of electrons is distorted
by energy losses due to the emission of Bremsstrahlung
in the detector planes before the electromagnetic calorimeter.
Consequently, the dielectron mass distribution shows a more pronounced tail
towards smaller masses. The description of this tail taking into account
Bremsstrahlung losses as well as the radiative tail due to the decay
{\ensuremath{\mathrm{J}/\psi \rightarrow \mathrm{e}^+ \mathrm{e}^- \gamma}} was
adjusted by using our dilepton-triggered \JPsiee\ sample~\cite{Zoccoli} which
was recorded in the same data taking period with the same detector set-up.
Fig.~\ref{fig:eesel}a displays the dielectron invariant mass distribution
after applying the selection criteria listed in Table~\ref{tab:presel}.
Using an exponential to describe the background we obtain
$57\pm13$ \JPsiee\ events (Table~\ref{tab:NumJPsi}). As for the muons, the
exponential background can be motivated by the like sign mass spectrum
shown in Fig.~\ref{fig:eesel}b.

\section{Efficiency Determination}
\label{MC}

A Monte Carlo simulation is used to determine the 
\JPsill\ efficiencies.
The Monte Carlo samples for $pA \rightarrow \JPsi + X$
are generated in two steps. First, a $c\overline{c}$ pair is generated
with {\sc Pythia} 5.7~\cite{sjo94} such that a \JPsi\ is always produced.  
The generated events are reweighted such that the resulting cross sections
conform to the parameterizations  
\begin{displaymath}
   \frac{d \sigma}{d \mt ^2  } \; \propto \; 
   \left[ 1 + \left( \frac{35 \cdot \pi \cdot \mt }{256 \; \cdot \langle p_\mathrm{T} \rangle } \right)^2
   \right] ^{-6}  
\end{displaymath}
and 
\begin{displaymath}
   \frac{d \sigma}{d \xeff  } \; \propto \; ( 1 - | \xeff | )^c \;.
\end{displaymath}
 
\noindent
The average transverse momentum of
$\langle p_\mathrm{T} \rangle = 1.29\pm 0.01$ GeV/c 
is taken from a preliminary \hb\ analysis of the \JPsi\ sample recorded
with a dilepton trigger~\cite{Zoccoli}. The slight increase of the
average transverse momentum with the mass of the nucleus is neglected.
We take $c= 6.38\pm 0.24$ as measured in
proton-silicon collisions at 800\,GeV/c~\cite{Alexopoulos}.
The \JPsi 's are generated without polarization. The influence of a 
possible \JPsi\ polarization within limits given by the experimental
results~[22-25] is taken into account as a systematic error. 
After the generation of the \JPsi\ the remaining energy is
given as input to {\sc Fritiof} 7.02~\cite{hpi92} which generates the underlying
event taking into account further interactions inside the nucleus.

The detector response is simulated with the {\sc Geant} 3.21
package~\cite{GEANT}.  Realistic detector efficiencies, readout noise
and dead channels are taken into account. The simulated events are
processed by the same reconstruction codes as the data.
The total efficiencies, including track reconstruction as well as the influence
of the selection cuts, are summarized in Table~\ref{tab:effic}.
The trigger efficiency for events containing a \JPsi\ meson is above
$99\%$.
 
\begin{table}[hbt]
\begin{center}
\begin{tabular}{|l|c|c|}
\hline
Target &  $\epsilon$(\mpmm  ) & $\epsilon$(\epem  )
\\ \hline \hline
C Below 1  & 0.337 $\pm$ 0.003 & 0.256 $\pm$ 0.003
\\ \hline
C Inner 2  & 0.339 $\pm$ 0.004 & 0.261 $\pm$ 0.004
\\ \hline
Ti Below 2 & 0.325 $\pm$ 0.003 & 0.247 $\pm$ 0.002
\\ \hline 
W Inner 1  & 0.317 $\pm$ 0.002 & 0.235 $\pm$ 0.002
\\ \hline
\end{tabular}
\vspace*{0.3cm}
\caption{Total efficiency of reconstruction and selection cuts in the
rapidity interval $-1.25<y<+0.35$. The uncertainties quoted are the
statistical uncertainties of the MC simulation. \label{tab:effic}}
\end{center}
\end{table}

\section{Results}
\label{Results}

\subsection{Combined cross sections per nucleus}

The first step in determining the production cross sections is to combine, for each
target material, the results of the two different final states. This is possible because
the two measurements are statistically independent and compatible. The weighted average
takes into account the statistical errors as well as those systematic errors which
depend on the lepton species (contributions 1 - 2 of Sec.~\ref{SYSERR}). 
The visible cross sections per nucleus, i.e. the cross sections measured
in the rapidity interval of $-1.25<y<+0.35$, are given by
\begin{displaymath}
   \Delta \sigma_{pA}^{J / \psi} =
   \frac{N_{i}}{Br(J / \psi \rightarrow \dilepton ) \cdot 
   \epsilon_{i} \cdot \mathcal{L}_{i}  }
\end{displaymath}
where 
\begin{itemize}
\item $N_{i}$, $\epsilon_{i}$ and $\mathcal{L}_{i}$ are the
measured \JPsi\ yield, the efficiency (Table~\ref{tab:effic})
and the integrated luminosity (Table~\ref{tab:numevents}) for a particular target,
\item $Br$ is the branching fraction which is $(5.88 \pm 0.10)~\%$ for \JPsimm\ and 
$(5.93 \pm 0.10)~\%$ for \JPsiee ~\cite{PDG}.
\end{itemize}

\noindent
The combined cross sections $\Delta \sigma_{pA}^{J / \psi}$,
summarized in Table~\ref{tab:results},
are almost independent of the assumptions made on the
differential distributions (see Sec.~\ref{MC}). However, for the total
cross sections an extrapolation factor $1/f$ must be applied which depends on the
shape of the rapidity distribution. This factor $f = 0.631 \pm 0.010$
is determined from the rapidity distribution of the \JPsi\ mesons generated
in the Monte Carlo simulation.

The total cross sections obtained by extrapolation of the visible cross sections
\begin{displaymath}
  \sigma_{pA}^{J / \psi} = \frac{\Delta \sigma_{pA}^{J / \psi}}{f}
\end{displaymath}
are also given in Table~\ref{tab:results}.

\begin{table}[bth]
\begin{center}
\begin{tabular}{|l|c|c|c|}
\hline
Target   &  A  & $\Delta \sigma_{pA}$ ($\mu$b) & $\sigma_{pA}$ ($\mu$b)
\\ \hline  \hline
carbon    &  12.011  & $ 4.0 \pm  0.7 \pm 0.3 $ & $  6.3 \pm  1.1 \pm 0.5 $
\\ \hline 
titanium  &  47.867  & $ 17. \pm  6.  \pm 1.  $ & $ 27.  \pm   9. \pm 2. $
\\ \hline 
tungsten  &  183.84  & $ 75. \pm  11. \pm 5.  $ & $ 118. \pm  18. \pm 8.  $
\\ \hline
\end{tabular}
\vspace*{0.3cm}
\caption{Visible $\Delta \sigma_{pA}$  (in the rapidity interval $-1.25<y<+0.35$)
and total $\sigma_{pA}$ \JPsi\ cross sections per nucleus of
atomic weight A. The errors quoted indicate the statistical and
systematic uncertainties, respectively.
\label{tab:results}}
\end{center}
\end{table}

\subsection{Systematic Uncertainties}
\label{SYSERR}

The total systematic uncertainty of the production cross sections is composed of
the following contributions:
\begin{enumerate}
\item
Particle identification cuts are varied over a wide range and, for each set of
cut values, the cross section is evaluated. The variation
of the cross section corresponds to $2.5 \%$ for \JPsimm\ and
$2.0 \%$ for \JPsiee , respectively. 
\item
Allowing for extreme variations of both signal shape and background parameterization,
we estimate the rms uncertainty on signal counting by dividing the difference of the
extreme values by $\sqrt{12}$ resulting in 2.5$\%$ for \JPsimm\ and 8$\%$ for \JPsiee .
\item
The luminosity uncertainties of each data sample are shown in Table~\ref{tab:numevents}.
\item
The overall scaling error of the luminosity is 2.0$\%$~\cite{lumi}.
\item
The track reconstruction efficiency is tested separately for VDS and OTR
using $K^0_s$ decays. One of the charged pions is reconstructed based on VDS (or OTR)
and RICH/ECAL information only. Applying the same procedure to Monte Carlo data,
we obtain an uncertainty of the track reconstruction efficiency of 1.5$\%$ per track.
\item
The uncertainty of the Monte Carlo production model within our kinematical range is 2.5$\%$.
This includes variations of the \mt\ shape as well as of the polarization parameter.
Varying the exponent $c$ of the \xeff\  distribution within errors has negligible
influence on our reconstruction efficiency. Reducing the exponent $c$ from 6.38 to 5.0
corresponds to a $1 \%$ change in the efficiency.
\item
Each of the branching fractions  for \JPsimm\ and \JPsiee\
has an uncertainty of 1.7$\%$~\cite{PDG}.
\item
The uncertainty on the extrapolation to the full rapidity range is determined by varying the
exponent of the \xeff\ parameterization within the errors to be 1.5$\%$.
\item 
Using $\sigma_{pA}^{J / \psi} =  \sigma_{pN}^{J / \psi} \cdot  A^{\alpha}$ as
parameterization of the A-dependence, the uncertainty of
$\alpha = 0.96 \pm 0.01$~\cite{Leitch} corresponds to a 3.6$\%$ uncertainty
of the cross section per nucleon.
\end{enumerate}
  
\subsection{Cross sections per nucleon}

\begin{figure}[htb]
\begin{center}
\epsfig{figure=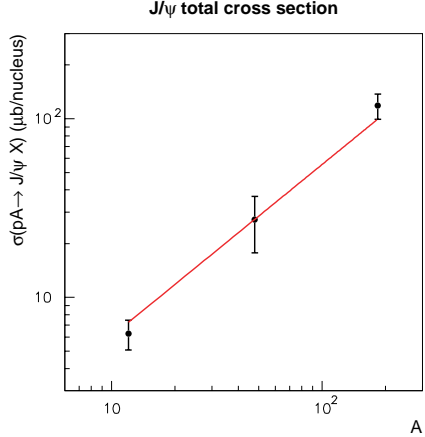,width=0.4\columnwidth}
\caption{The total \JPsi\ cross section on nuclear targets C, Ti and W.
The error bars indicate the statistical and systematic uncertainties 1 - 3
explained in Sec.~\ref{SYSERR}. The line shows the result of the fit described in
the text using $\alpha = 0.96$. 
\label{fig:JPsiadep}}
\end{center}
\end{figure}

As a last step, the results of the three data samples are combined to
determine the cross section per nucleon $\sigma_{pN}^{J / \psi}$.
The cross section for \JPsi\ production
on a nuclear target of atomic weight A is parameterized as  
\begin{displaymath}
   \sigma_{pA}^{J / \psi} =  \sigma_{pN}^{J / \psi} \cdot  A^{\alpha}.
\end{displaymath}
By fitting this expression to our data we can determine the cross section per
nucleon. However, the limited statistics of our sample does not allow for a
precise measurement of the A-dependence. Therefore we fix the parameter
to the mid-rapidity measurement $\alpha = 0.96 \pm 0.01$ of E866~\cite{Leitch}.
Fig.~\ref{fig:JPsiadep} demonstrates that our result is in good agreement
with this assumption ($\chi ^2 / NDF = 1.5/2 $).
For this fit only the statistical and the relevant systematic uncertainties
(1 - 3 of Sec.~\ref{SYSERR}) are taken into account.

As a result we obtain

\begin{displaymath}    
   \sigma_{pN}^{J / \psi} =  663 \pm 74 \pm 46 \; \mathrm{nb/nucleon}
\end{displaymath}
for the total \JPsi\ production cross section per nucleon.
The systematic uncertainty quoted comprises all contributions described
in Sec.~\ref{SYSERR}.
 
The mid-rapidity cross section can be derived by applying a factor
$\eta = 1.50 \pm 0.03$ which corrects for the difference between the differential
cross section at $y=0$ and the visible cross section, which is an average of the
interval $-1.25<y<+0.35$ with $\Delta y = 1.6$  
\begin{displaymath}
  \left. \frac{d\sigma_{pN}^{J / \psi}}{dy}\right|_{y=0} 
 = \frac{\Delta \sigma_{pN}^{J / \psi}}{\Delta y} \cdot \eta  
 =  392 \pm 44 \pm 27 \; \mathrm{nb/nucleon.}
\end{displaymath}

\section{Comparison with other results}

\begin{figure}[htb]
\begin{center}
\epsfig{figure=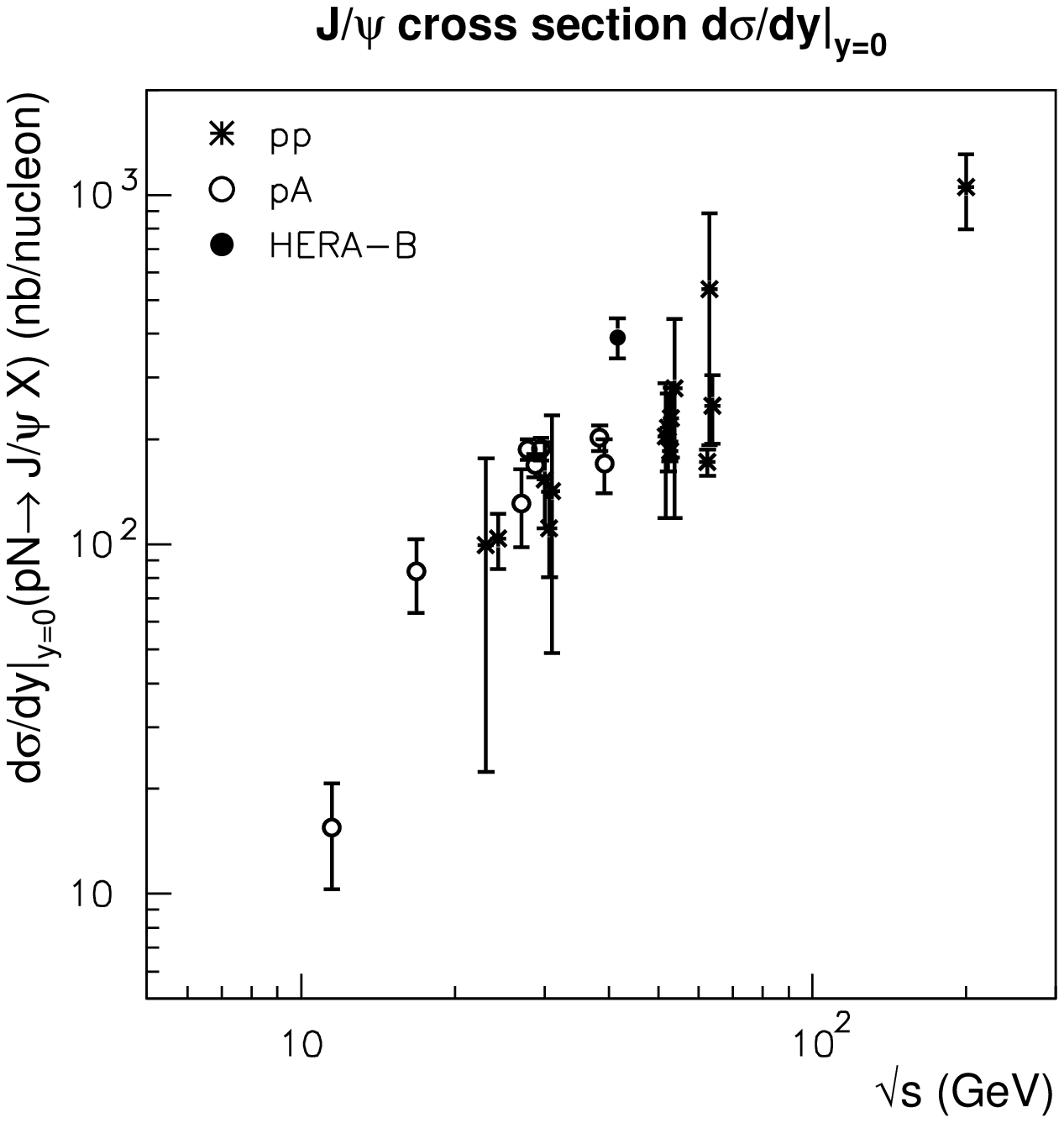,width=0.5\columnwidth}
\epsfig{figure=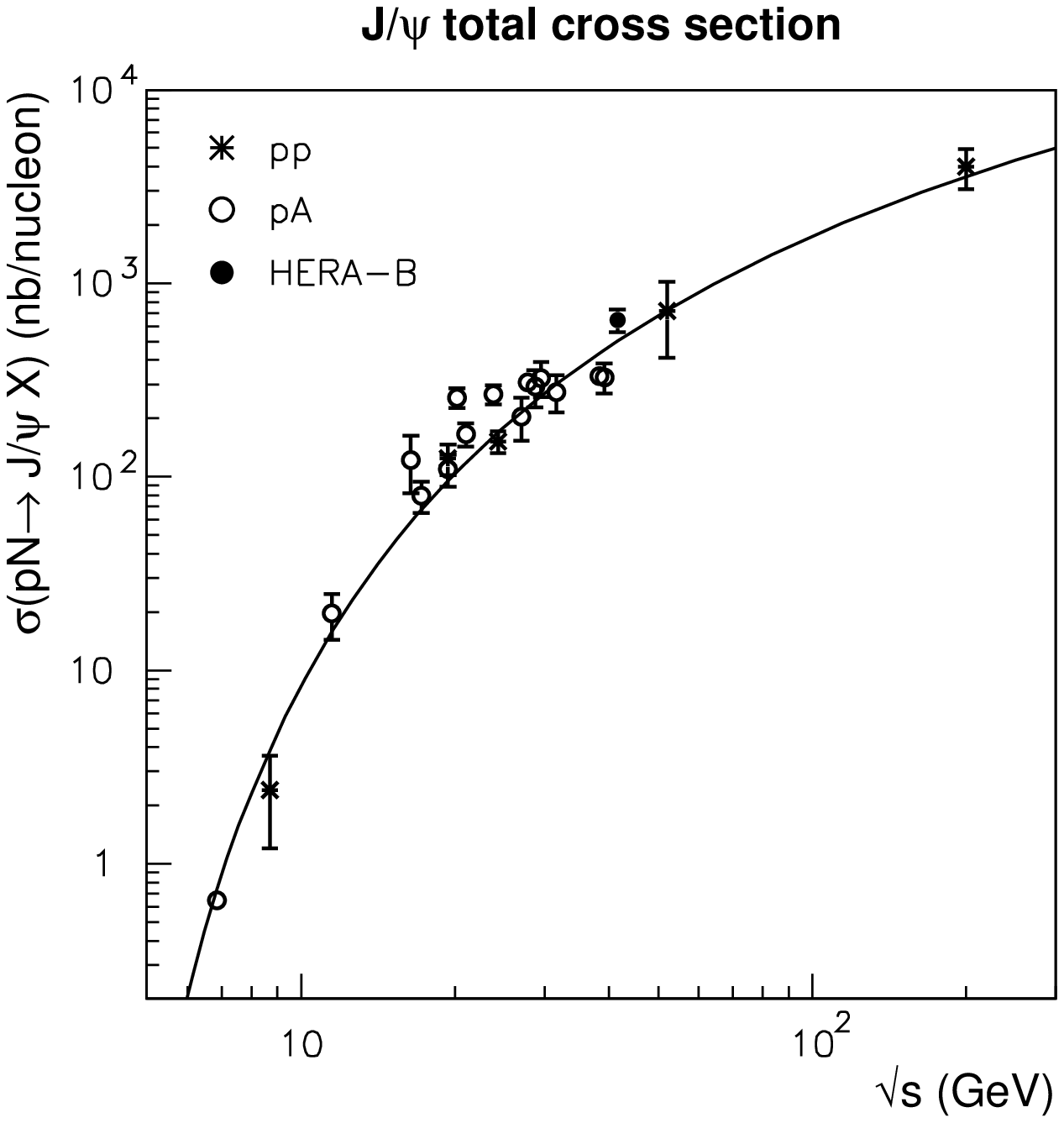,width=0.5\columnwidth}
\caption{\JPsi\ production cross sections in proton-induced interactions
[22-24, 30-47].
pp and pA measurements are indicated by different symbols.
a) Differential cross section $d\sigma_{pN}/dy$ at $y$=0; 
b) total cross section $\sigma_{pN}$ together with the fit described
in the text.
\label{fig:sqsxsect}}
\end{center}
\end{figure}

Comparing our results to the measurements of previous experiments in
proton-induced interactions requires correcting all results for the
same branching fractions and A-dependence. All results shown in
Fig.~\ref{fig:sqsxsect} are updated for the latest values of the
branching fractions quoted in Sec.~\ref{Results} and a target mass
dependence assuming $\alpha = 0.96 \pm 0.01$. The systematic
uncertainties are recalculated accordingly.

It is obvious that the experimental results are far from being consistent.
Cross sections measured at nearby beam energies are not compatible when taking
into account just the errors quoted by the experiments. 
Also shown on the figure is a fit to the published data on proton-induced
\JPsi\ and \Psizs\ production in the context of a next-to-leading order NRQCD
calculation~\cite{Maltoni}. Several fits are performed with different subsets
of measurements and different parton distribution functions (PDF).
The best description of the data is obtained with the PDF
MRST2002~\cite{MRST2002}. 

At \sqs\ =41.6~GeV, the fit gives a cross section  of
\begin{displaymath}    
   \sigma_{\mathrm{average}}^{J / \psi} =  502 \pm 44 \; \mathrm{nb/nucleon}
\end{displaymath}
which is in reasonable agreement with the result presented here. The error
takes into account both fit uncertainties as well as systematics due to
the parton distribution function.
Details of the procedure together with tables of all experimental input
used are the subject of a separate paper~\cite{Maltoni2}.

\section{Summary}

A data sample taken with a minimum bias trigger by the \hb\ experiment in interactions of
920 GeV/c protons with carbon, titanium and tungsten targets has been used 
to determine the production cross section of \JPsi\ mesons. In our
data sample we find  $100\pm12$ \JPsimm\ and $57\pm13$ \JPsiee\
candidates. After correcting for the efficiency of the selection criteria
within the range in rapidity of $-1.25<y<+0.35$ and combining both final
states, we obtain the visible cross sections per nucleus within the detector
acceptance. Using 
$\sigma_{pA}^{J / \psi} =  \sigma_{pN}^{J / \psi} \cdot  A^{0.96 \pm 0.01}$
to parameterize the A-dependence, and
extrapolating this measurement to the full $y$ range, the total
production cross section per nucleon is 
\begin{displaymath}
   \sigma_{pN}^{J / \psi} =  663 \pm 74 \pm 46 \; \mathrm{nb/nucleon.}
\end{displaymath} 
For the cross section at mid-rapidity we obtain
\begin{displaymath}
  \left. \frac{d\sigma_{pN}^{J / \psi}}{dy}\right|_{y=0} =  392 \pm 44 \pm 27 \; \mathrm{nb/nucleon.} 
\end{displaymath}  

An NRQCD based fit of published results on \JPsi\ and \Psizs\ production predicts 
$\sigma_{\mathrm{average}}^{J / \psi} =  502 \pm 44 \; \mathrm{nb/nucleon}$
at 920 GeV/c, in reasonable agreement with this measurement.

\section*{Acknowledgments}

We are grateful to the DESY laboratory and to the DESY accelerator
group for their strong support since the conception of the HERA-B
experiment.  The HERA-B experiment would not have been possible
without the enormous effort and commitment of our technical and
administrative staff.

\end{document}